# Improving ECG Classification Interpretability using Saliency Maps

Ms Yola Jones
Robertson Centre for Biostatistics
University of Glasgow
Glasgow, Scotland
y.jones.1@research.gla.ac.uk

Dr Fani Deligianni
School of Computing Science
University of Glasgow
Glasgow, Scotland
Fani.Deligianni@glasgow.ac.uk

Dr Jeff Dalton
School of Computing Science
University of Glasgow
Glasgow, Scotland
Jeff.Dalton@glasgow.ac.uk

*Abstract*—Cardiovascular disease is a large worldwide healthcare issue; symptoms often present suddenly with minimal warning. The electrocardiogram (ECG) is a fast, simple and reliable method of evaluating the health of the heart, by measuring electrical activity recorded through electrodes placed on the skin. ECGs often need to be analyzed by a cardiologist, taking time which could be spent on improving patient care and outcomes.

Because of this, automatic ECG classification systems using machine learning have been proposed, which can learn complex interactions between ECG features and use this to detect abnormalities. However, algorithms built for this purpose often fail to generalize well to unseen data, reporting initially impressive results which drop dramatically when applied to new environments. Additionally, machine learning algorithms suffer a 'black-box' issue, in which it is difficult to determine how a decision has been made. This is vital for applications in healthcare, as clinicians need to be able to verify the process of evaluation in order to trust the algorithm.

This paper proposes a method for visualizing model decisions across each class in the MIT-BIH arrhythmia dataset, using adapted saliency maps averaged across complete classes to determine what patterns are being learned. We do this by building two algorithms based on state-of-the-art models. This paper highlights how these maps can be used to find problems in the model which could be affecting generalizability and model performance. Comparing saliency maps across complete classes gives an overall impression of confounding variables or other biases in the model, unlike what would be highlighted when comparing saliency maps on an ECG-by-ECG basis.

## I. Introduction

In 2016, approximately 1 in 3 deaths in the US were from cardiovascular disease, with heart disease being the number one cause of death today [1]. The electrocardiogram (ECG) is still the most accessible and inexpensive tool for diagnosing these conditions [2]. Despite its frequent use, the rates of misdiagnosis from ECGs are still high. A study in 2001 found that 33% of the 300 million ECGs performed annually had some interpretation discrepancy, with approximately 11% of these discrepancies resulting in a change in patient care [3] [4]. While this study is not recent, it does highlight the impact ECG misinterpretation can have, and the number of ECGs being recorded annually. ECG interpretation algorithms therefore need to be accessible, accurate, and understandable so errors can be quickly found and addressed before they impact patient care.

In light of this, ECG classification is commonly addressed in machine learning. However, models created often fail to identify and handle clinical concerns which limit the uptake of these models in patient-facing applications. Models often have high accuracy, sensitivity and specificity, but fail to generalize well to unseen data, for example new patients or new environments. They also do not address the interpretability gap which limit the use of these algorithms in patient-facing healthcare, due to the 'black-box' nature of ML in which clinicians do not trust a model which cannot explain its decisions, especially when these decisions can affect patient care and outcomes.

In this paper, we aim to address some common concerns, by building two model architectures based on state-of-the-art work in this area, namely a convolutional neural network (CNN) and a long short-term memory network (LSTM).

We train these models using the MIT-BIH arrhythmia dataset [5] split into single beats, where each beat is classified into one of eight beat classifications. The results of these models are compared, and we address the 'black-box' interpretability issue, in which models offer little insight into how a decision has been made. This is done by analyzing class-by-class results and highlighting common mistakes using saliency maps. We also analyse what the model is focusing on for each ECG when deciding on a classification. This is done by segmenting saliency maps, repeated for each ECG in our dataset.

Our method allows us to quantitatively compare what features of the ECG are being focused on by the machine learning model for each of the classes. It gives us an impression of what patterns are being learned, highlighting confounding variables or biases affecting the model across complete classes, which could be affecting generalizability and overall performance.

## II. Background

### A. ECG Morphology

The electrocardiogram (ECG) is a non-invasive method of analyzing the heart, measuring electrical activity from multiple angles by placing electrodes at different points on the skin. The key features are the P-wave, QRS complex and T-wave, each representing a different stage of the heartbeat, and which can be used to help detect abnormalities by relating them with the underlying phase of the cardiac cycle. ECGs are highly susceptible to noise, both high and low frequency, often attributable to improper electrode placement, baseline wander, and muscle movement [6]. These effects often obscure key features necessary to determine the health of the heart, which can make these ECGs difficult to classify.

### B. Comparison with other models

In 2019, Hannun et al [8] published a paper classifying single-lead ECGs from 53,549 patients with the Zio monitor from iRhythm technologies. Their network consisted of 16 residual blocks with two convolution layers per block, and utilized shortcut connections, batch normalization, and dropout with a final fully-connected softmax layer classifying each sample into one of 12 classes, including normal sinus rhythm and noise. They trained and tested their model using ECGs from the Zio monitor, but also tested their model on the 2017 PhysioNet Challenge dataset to demonstrate its generalizability, with impressive results.

Following on from this work, Ribeiro et al [9] recently built a dataset of ECGs from 1,676,384 patients, training a model to detect six abnormalities. They employed the same model structure as [8], but found best results when using a simplified network with roughly a quarter of the layers. They use four residual blocks, each block consists of convolution layers, dropout, batch normalization and shortcut connections.

Pandey et al in 2019 [10] used an 11-layer CNN to classify beats in the MIT-BIH arrhythmia dataset into five classes, using Synthetic Minority Oversampling TEchnique (SMOTE) to handle the dataset imbalance shown in Figure 1. Their network included four convolution and max pooling blocks, followed by two fully connected ReLU layers and a final fully connected softmax layer to classify beats into the five classes. The model was tested using the hold-out method, by randomly splitting their beats into training and testing datasets.

Focal loss is an extension of the cross-entropy loss function originally proposed by [11] which makes the model focus on difficult to classify samples, thereby attempting to address the class imbalance issue commonly found in healthcare datasets. Studies do show it causes modest improvement compared to the traditional cross-entropy loss function, and compared to using cross-entropy with resampling to account for imbalanced datasets [12]. In 2019, Gao et al [13] proposed the use of an LSTM to classify beats in the MIT-BIH arrhythmia dataset using this loss function. They used a 64-node LSTM layer followed by two fully-connected layers to classify beats into one of eight classes and trained their model for 350 epochs. They tested their model using the hold-out method, in which 90% of their data was used for training and 10% for testing. The authors found using the focal-loss as opposed to cross entropy loss function improved results and found improvement when first denoising their ECGs using the Daubechies 6 (db6) discrete wavelet transform.

*C. Machine Learning in Clinical Practice*

Despite the promising results of these and similar models, there is still a very small number of ML algorithms being used in healthcare. Healthcare is a challenging environment to model, due to intrinsic difficulties unique to this domain. One such issue is that patient population characteristics are diverse and evolving, and can vary widely between different patient groups [14]. Healthcare datasets often also include confounding variables which are difficult to spot without extensive examination of the data [15]. In addition, healthcare datasets are often very imbalanced, due to the influence of rare conditions and the fact that people are generally more likely to be healthy than sick.

Ahmad et al. in [16] highlight the need for interpretable models. When applying ML to healthcare, life-or-death decisions could be being made, and therefore the 'black-box' nature of ML needs to be addressed. This will allow clinicians to understand and verify why a decision has been made in order to address intrinsic biases or effects due to confounding variables ML is often affected by.

Several techniques have been proposed as a solution [17], especially for convolutional neural networks, including occlusion, filter visualization, and class activation mapping [18]. Saliency maps are one such technique, producing a heatmap of how different areas of an image affect the final classification [19].

The idea of saliency maps is very simple, take an example ECG, feed it through the model to update the model gradients, then go back to the input layer to calculate the derivative of the output layer with respect to each pixel in the input image.

Strodthoff and Strodthoff in [20] apply DeepExplain [21] to models trained using the PTB diagnostic ECG database, comparing results to clinical interpretation of these ECGs. Vijayarangan et al. [22] use saliency maps and class-activation mappings to visualize both LSTM and CNN models using the MIT-BIH arrhythmia dataset. Both papers work towards uncovering underlying decisions made by these models, attempting to explain what motivates particular decisions. However, both works address interpretation on an ECG-by-ECG basis, failing to capture the overall rules or features a model has picked up across the entire dataset under test. This is a subtle but important difference, while interpreting each ECG individually is useful, it is difficult to uncover systematic biases or confounding variables when models are investigated with individual ECGs. A way of capturing overarching rules and patterns is needed to fully investigate model behavior across multiple ECGs, which can then be used to explain or predict any issues in model generalizability to unseen data.

### III. METHODS

For this study, we used supervised learning to investigate two model architectures, a convolutional neural network (CNN), and a long short-term memory network (LSTM). We based our CNN on the architecture proposed by Ribeiro et al in [9], and our LSTM on work done by Gao et al in [13] in order to compare our results to existing work, and tuned hyperparameters to optimize results.

We trained and tested these models using the MIT-BIH arrhythmia dataset, a database of 30-minute recordings of two-channel ambulatory ECGs, in which each beat is annotated by a cardiologist. This dataset is a standard dataset used for ECG classification tasks, and so provides a benchmark to which we can compare models. We performed beat extraction to segment the recordings into beats, each labelled with a particular class. The MIT-BIH arrhythmia dataset includes the location of the R-peaks with the annotations, therefore we define a single beat as an ECG segment centered around one single R-peak, zero-padded to match the length of the longest sample in our dataset. The goal of our models was to correctly classify each beat into its beat classification.

After separating by beat, we trained and tested each of these models in two different ways. Firstly, we use a traditional hold-out method, where 65% of the beats in the dataset were used for training, 10% for validation, and 25% for testing. Since patient characteristics can influence ECG morphology, a normal beat from a patient will likely look very similar to another normal beat from the same patient, which means testing in this way without separating by patients first could result in data leakage. Data leakage is a common machine learning mistake in which the model reports erroneously high results due to the model being tested on data it has already seen [23], i.e. when data from the same patient is used in both the training and testing dataset. We attempt to address this using the second method, by using different patients for the training and testing datasets. 43 of the 48 recordings are as the training and validation dataset, and the remaining five patients as testing.

### A. MIT-BIH Dataset

The MIT-BIH arrhythmia dataset was selected for this study as it is a large open-source dataset which has been used to train many models for ECG classification, allowing us to compare our results with previous work. The structure of this dataset has been well documented in previous studies and contains 48 30-minute recordings of two-channel ambulatory ECGs from 47 patients sampled at 360 Hz, recorded between 1975 and 1979. Each beat in the dataset has been annotated by a cardiologist and labelled as one of 19 classes as documented on the PhysioNet website [24]. Since each beat is labelled at the R-peak, we used this to extract our beats, a method that has been shown to be 99% accurate [13]. Beat extraction is performed by halving the distance between beats on either side of each R-peak shown in Figure 2.

After removing non-beat annotations, the class distribution for this dataset is given in Figure 1. We excluded classes S and J due to lack of samples, to create 8 different beat classifications.

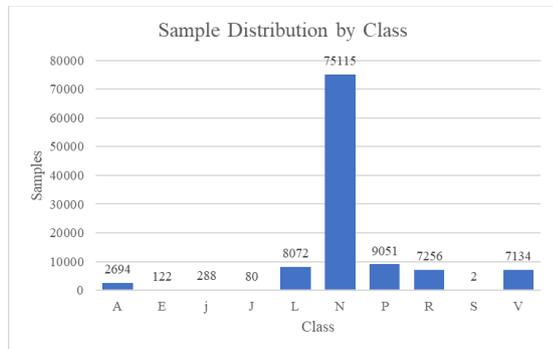

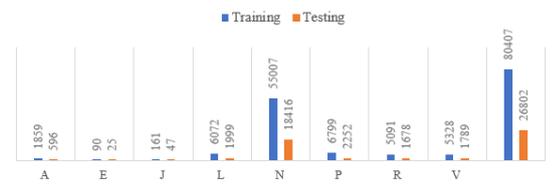

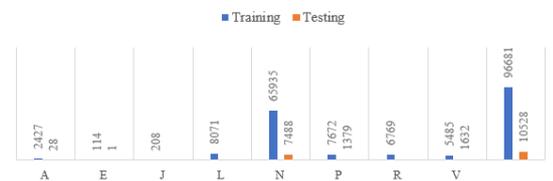

Figure 1: MIT-BIH Arrhythmia Dataset Class Distribution

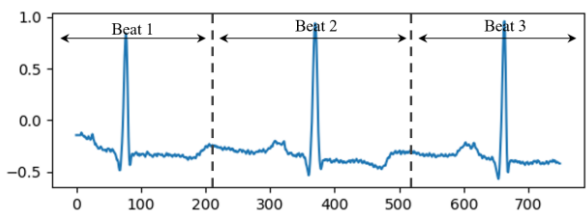

Figure 2: Beat Extraction

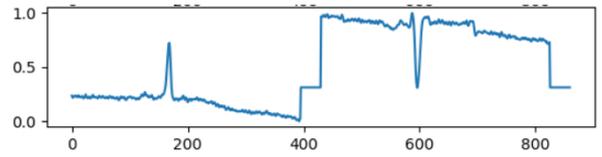

Figure 3: Complete Beat Example

### B. Dataset Creation

For simplicity, we trained models which rely on consistent input length for each beat. To do this, we planned to zero-pad each beat to the length of the longest beat. However, since the longest beat was over 1200 samples long, and only 2% of beats were longer than 430 samples, we decided to exclude this top 2% from our study in order to reduce training time due to overly long beats filled with mostly redundant information.

After building some initial models, we found that including both leads in model training improved performance. For the final dataset, we therefore zero-padded each lead to 430 samples, appending the second lead to the first to create one 860-length beat, as shown in Figure 3 .

This created a total dataset of 107 209 beats in 8 different classes for model training, validation and testing.

Since the class distribution is so uneven (Figure 1), and normal beats include at least 8 times more samples than each of the other classes, we downsampled the number of beats in the normal class to match the second-largest class, upsampling the rest of the classes to create an even distribution of beats across all eight classes. Downsampling was performed by randomly selecting samples to delete from the majority class, using sklearn's resample method, and upsample was performed by randomly selecting a population of unique samples to be duplicated, and duplicating these until the desired class size is reached.

Finally, we normalized each ECG to between zero and one before training, validation and testing.

### C. Testing

For the traditional hold-out method, we randomly selected 25% of beats to become the testing dataset. For testing using unseen patients, we followed the same pre-processing method described above, but first separating the beats by patient. When removing patients from the training dataset we run the risk of removing too many samples from one or two classes for training, and as a result create a training dataset which cannot recognize these classes due to insufficient samples. In order to combat this, we experimented with randomly removing patients from the testing dataset, to ensure each class had enough samples while not introducing bias by having the testing dataset comprised only of classes we know to be easy to distinguish.

Using this process, we found that using patients 104, 113, 119, 208 and 210 from the MIT-BIH arrhythmia dataset gave us a wide enough array of classes to test, while not impacting the class distribution of the training dataset too greatly. The class distributions of the training and testing sets for both methods are shown in Figure 1.

### D. Saliency Map Visualisation

An ECG can be thought of as a 1D image, and therefore this saliency maps can be used to visualize what areas of an ECG

influence model classification. To inspect our models using this method, we took the model state at the final epoch of the trained model and found the saliency map of an input ECG using keras' visualize_saliency function. This function is used to visualize the CNN trained in this study and is pointed at the final convolutional layer of the model to get a map of the areas of the ECG which influence the filter, plotted along the input ECG to highlight each area.

From discussion with clinicians, we would expect to see the model highlighting key morphological features of the ECG, such as the P-wave (or area where the P-wave should be if there is no P-wave present), QRS complex, and T-wave. It could also highlight areas we do not expect which could be investigated further, from the ability of these techniques to pick up complex features and interactions between pixels in an input image.

We can also use this method to investigate if the model is picking up areas we do not want, for example noise. In a dataset such as this where a lot of data is taken from few patients, it is very possible that confounding factors such as lead placement or fast heart rate could influence classification. This would hopefully be reveled through further investigation of the saliency maps.

*E. Saliency Map Segment Analysis*

After having generated the saliency plots using the method described above, we split each ECG into 0.1 second segments, starting at the R-interval in each lead and working outwards. We then took the mean saliency value for each of these blocks, in order to get a single saliency value per segment for each ECG.

Performing this saliency-segmentation for each ECG within each class allows us to find the 'median' saliency-segmentation plot for each class. This allows us to quantitatively compare where the model is looking within each class and allows us to make inferences about what the model is looking for. It also allows us to easily find an investigate outliers, samples which are likely to be misclassified by the model.

## IV. RESULTS

We trained all models on a laptop with a dual core i7 processor with an NVIDIA GeForce 940MX graphics card. We used Python version 3.7.3, TensorFlow version 2.1.0, and Keras version 2.2.4-tf to train and test our models. All code can be found at the GitHub link https://github.com/jonesy30/ECGClassification.

After some preliminary analysis, we found that denoising did not improve results by any large margin and so did not include this step in data pre-processing.

The original paper proposed by [13] used 350 epochs, however we found that our model converged quickly and so stopped training after 40 epochs. We use the Nadam optimizer with an initial learning rate of 0.001 as in the original paper. For our CNN, we used the Adam optimizer with an initial learning rate of 0.001, decreasing the learning rate by a factor of 10 when results plateau as in the model proposed by Ribeiro et al. in [9]. We use a batch size of 128 for both models.

Complete results for the accuracy, F1 score (macro), precision, recall and specificity for each model with each testing method can be seen in Table 1. The CNN seems to outperform the LSTM, with a dramatic drop in results in the leave-patients-out testing method compared to the hold-out method, highlighting that the models are likely struggling to generalize to unseen patients.

The confusion matrix for each model is shown in Figure 4. In the confusion matrix, the y-axis shows the true labels, the x-axis shows classes predicted by the model, where each row gives the proportion of items in the true class that were predicted as being in each class.

For example, in the top row of the CNN hold-out confusion matrix, 97% of APB samples were predicted by the model as being APB, 3% of APB samples were predicted as normal. A nan value means there are no samples in that class to be tested.

While testing using the hold-out method is the simplest of techniques, it can result in some severe flaws in reported results due to data leakage. The MIT-BIH dataset consists of 30-minute samples from 47 patients. Splitting these recordings by beats and randomly selecting a subset of these beats for testing means that in almost all cases, patients will have beats in both the training and testing dataset. Since patients will have similar characteristics (electrode placement, distance between heart and electrode, comorbidities, medications, etc.), there is likely to be common morphology between beats, a normal beat taken from a patient will likely look similar to a normal beat taken from the same patient. This introduces data bleeding between the training and dataset, raising questions into the apparent success of models tested in this way.

Using the hold-out method, as with the original papers, our models are comparable to state-of-the-art (shown in Table 2 and

Table 3), with most metrics within 5% of state-of-the-art, although further work needs done to match these models in all metrics. However, the models do not perform well when trained and tested using unseen patients, dropping between 40 – 60% for precision, recall and macro F1 compared to the same model trained with the hold-out method. This could be the result of class imbalance, if leaving a patient out of training causes few samples of a particular class to be present in the training set. However, it is more likely an issue of generalizability, the models do not perform well on unseen data.

We include a per-class breakdown for each of the four models in Figure 6. The patients who were included in the leave-patients-out testing set had no beats in the LBBB, RBBB, ventricular escape or junctional escape beat classes, so these are more challenging to comment on.

However, it is clear for all tests that the CNN model tested using the hold-out method performs well with normal, paced, and premature ventricular contraction beats. Atrial premature beats perform well using the hold-out method but poorly when testing with unseen patients, despite the class distribution being maintained between the hold-out and leave-patients-out testing method. LBBB and RBBB seem to perform well, with junctional and ventricular escape beats performing poorly, although we cannot comment on their generalizability to unseen data.

Paced beats consistently perform well in all models, even on unseen data. This makes sense, as paced beats look very different compared to other types of beats. Usually, the original impulse which starts the heartbeat comes from a part of the heart called the sinoatrial node, providing an electrical impulse to initiate the heartbeat. Paced beats are beats which are initiated by a pacemaker, a foreign device implanted into the heart to regulate rhythm. Since the heartbeat is initiated by a foreign device as opposed to a natural part of the heart, paced beats have a different morphology in the QRS complex compared to other heartbeats, making them very easy to spot. This unique morphology is likely to have been recognized by the model, hence its generalizability to unseen patients.

In contrast, junctional escape beats consistently perform the worst in all models. An explanation for this could be the lack of samples in the training set, although interestingly it is not the smallest category. Ventricular escape beats consistently perform much better, despite having 93 less samples.

### A. Saliency Maps

Figure 5 gives an example of a raw saliency map which are used to visualize the final convolutional layer of our CNN model in order to determine where the model is looking when it makes its classification. The plots show the raw saliency value (the plot colors) on top of the ECG, so we can see exactly what areas of the ECG are being highlighted by the model. Saliency color values are shown in the color bar on the right of the plots, with purple representing areas which are mostly ignored by the model, and green and yellow representing areas which largely contribute to the final classification.

We use the raw plots of the beats in our dataset to generate our saliency-segment values for each class in Figure 7. The plots in this figure represent the median saliency map across samples for each class in the CNN hold-out model, with each block representing the median saliency value for every 0.1 second segment. The same scale has been used for these plots as with the raw plots (Figure 5), and so larger numbers (tending towards 1) are dampened by taking the median of these segments. Therefore, shown in Figure 7, yellow and green represent blocks that contribute more to the final classification decided by the model, and red indicating blocks that are mostly ignored.

As can be seen, the models tend to pay most attention to the QRS complex in the first lead, although by how much does depend on the class. Surprisingly, the QRS complex for the ventricular escape beat is ignored in the first lead, but highlighted heavily in the second beat, which could be an indication of noise as a confounding factor. Most samples in this class come from the same patient in the MIT-BIH dataset, which could support this claim. In both the normal and atrial premature beats, the saliency-by-segment plot is very similar. In these classes, the T-wave in both leads is ignored, only highlighting the QRS complex in the first lead and the end of the second lead (possibly to determine the length of the heartbeat). This similarity of saliency plots, specifically ignoring key features, could explain the poor performance when testing using unseen patients.

In contrast, investigating the results from our saliency-segment plots (Figure 7) in a class that performs consistently well, i.e. paced beats, we can see that the model is focusing most of its attention on the R-peaks in both leads. This suggests that it has picked up the unique morphology of the QRS complex when looking at paced beats.

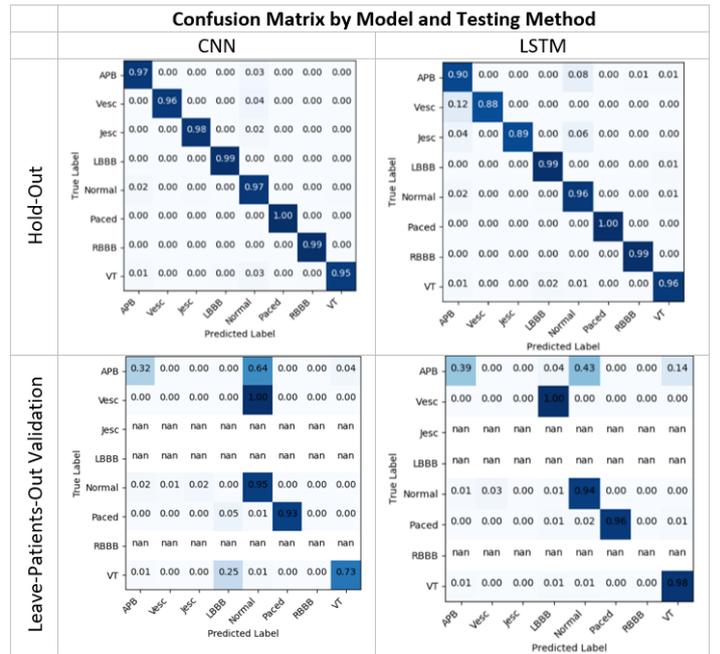

*Figure 4: Confusion Matrix for Each Class per Model per Testing Method*

TABLE 1: MODEL RESULTS PER METRIC

| Testing Method | | Model | |
|---|---|---|---|
| | | CNN | LSTM |
| Hold-out | Accuracy | 0.9710 | 0.9674 |
| | F1 Score | 0.8510 | 0.8422 |
| | Precision | 0.7924 | 0.7865 |
| | Recall | 0.9753 | 0.9478 |
| | Specificity | 0.9955 | 0.9951 |
| Leave-Patients-Out | Accuracy | 0.9094 | 0.9491 |
| | F1 Score | 0.3593 | 0.3856 |
| | Precision | 0.3805 | 0.3843 |
| | Recall | 0.3661 | 0.4089 |
| | Specificity | 0.9872 | 0.9923 |

TABLE 2: CNN COMPARISON TO STATE-OF-THE-ART

| Metric | CNN Model | |
|---|---|---|
| | Our Model | Pandey et al [10] (using 30% hold-out) |
| Accuracy | 0.971 | 0.983 |
| F1 | 0.851 | 0.899 |
| Precision | 0.792 | 0.861 |
| Recall | 0.975 | 0.955 |

TABLE 3: LSTM COMPARISON TO STATE-OF-THE-ART

| Metric | LSTM Model | |
|---|---|---|
| | Our Model | Gao et al [13] |
| Accuracy | 0.967 | 0.992 |
| F1 | 0.842 | 0.993 |
| Precision | 0.787 | 0.993 |
| Recall | 0.948 | 0.993 |
| Specificity | 0.995 | 0.991 |

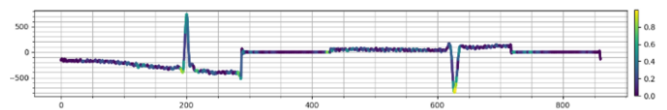

*Figure 5: Raw Saliency Map Example*

| | CNN (hold-out) | | | | | | | |
|---|---|---|---|---|---|---|---|---|
| | APB (A) | Vesc (E) | Jesc (j) | LBBB (L) | Normal (N) | Paced (P) | RBBB (R) | VT (V) |
| F1 Score | 0.75 | 0.67 | 0.51 | 0.98 | 0.98 | 1.0 | 0.98 | 0.95 |
| Precision | 0.61 | 0.51 | 0.34 | 0.96 | 1.0 | 0.99 | 0.97 | 0.95 |
| Recall | 0.97 | 0.96 | 0.98 | 0.99 | 0.97 | 1.0 | 0.99 | 0.95 |
| | LSTM (hold-out) | | | | | | | |
| F1 Score | 0.69 | 0.65 | 0.53 | 0.96 | 0.98 | 1.0 | 0.99 | 0.95 |
| Precision | 0.55 | 0.51 | 0.38 | 0.94 | 1.0 | 1.0 | 0.98 | 0.94 |
| Recall | 0.9 | 0.88 | 0.89 | 0.99 | 0.96 | 1.0 | 0.99 | 0.96 |
| | CNN (leave patients out) | | | | | | | |
| F1 Score | 0.1 | N/A | N/A | N/A | 0.97 | 0.97 | N/A | 0.84 |
| Precision | 0.06 | N/A | N/A | N/A | 0.99 | 1.0 | N/A | 0.99 |
| Recall | 0.32 | N/A | N/A | N/A | 0.95 | 0.93 | N/A | 0.73 |
| | LSTM (leave patients out) | | | | | | | |
| F1 Score | 0.16 | N/A | N/A | N/A | 0.97 | 0.98 | N/A | 0.98 |
| Precision | 0.1 | N/A | N/A | N/A | 0.99 | 1.0 | N/A | 0.98 |
| Recall | 0.38 | N/A | N/A | N/A | 0.94 | 0.96 | N/A | 0.98 |

*Figure 6: Per-Class Breakdown of Model Results*

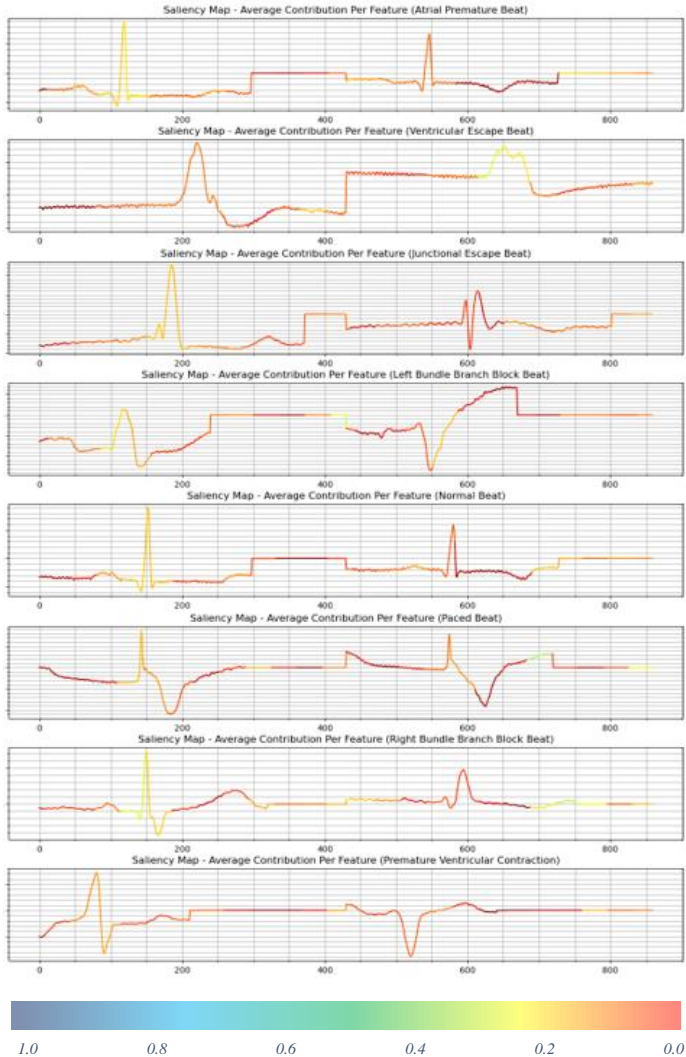

*Figure 7: Per-Segment Saliency Maps*

### B. Saliency-Map Differences Between Correctly Classified and Incorrectly Classified Samples

Using the per-segment plot for each class, we can find and investigate outliers, beats which are likely to be misclassified by the model. In the following discussion, we compare the median saliency value in each segment for incorrectly classified vs correctly classified beats in classes with the poorest performance, namely atrial premature beats, ventricular escape beats and junctional escape beats.

Figure 8 shows the results of these plots and are explained in more depth in the following section. Each plot is split into two leads, lead 1 on the left and lead 2 on the right. Each lead includes the R peak in the middle, labelled on the x-axis, with each point in the plot representing a 0.1 second segment block, moving outwards from the R-peak. The solid lines give the median values, and the shaded region gives the interquartile range for each of the points.

Each figure is broken into two, the top plot compares the correctly identified classes to the incorrectly identified classes, which is compared to the saliency-segment plots for all samples in that class (training and testing set combined). The second plot in each figure compares the incorrectly identified class to the class it was misclassified as.

*1) Atrial Premature Beats*

Figure 8 compares the saliency map plots from correctly classified beats with incorrectly classified beats and the total average for all beats in this class.

As can be seen, the incorrectly classified beats do differ from the correctly classified beats, the correct beats follow a much similar curve to the total. The second plot in the atrial premature beat plot of Figure 8 shows the incorrectly classified beats compared to the saliency-map segmentation plots of normal beats, which seem to fit the plot much better. When atrial premature beats are misclassified, they are often misclassified as normal, shown in Figure 4, which can be seen by the way the model interprets each ECG through the saliency maps.

*2) Junctional Escape Beats*

Unfortunately, due to the small number of junctional escape beat samples in the validation set, a small number of incorrectly classified samples results in a large error margin. In this group, 24 of the 25 samples were classified correctly, the 4% misclassification resulting from one incorrectly classified sample.

However, we can still compare the incorrectly classified ECG to the correct and total classifications, shown in Figure 8. As can be seen, the incorrectly classified beat doesn't match the typical saliency map for this class at all. The second plot in this figure compares the saliency-segment plot to the normal segment block. These plots are slightly more similar, and could explain this ECG being classified as normal, although doesn't match perfectly. It is likely that the model simply didn't know what class this ECG belonged to, and therefore categorized it as normal.

*3) Ventricular Escape Beats*

Likewise, with junctional escape beats, only one ECG was misclassified in this class, but this caused a larger error due to the small occurrence of samples of this class in the validation set. Also similar to junctional escape beat, this incorrectly classified beat seems to be being analyzed differently compared to the rest of the beats in this class. However, unlike the junctional escape beat, the incorrect and normal plots match each other very closely, giving an indication of why this beat was incorrectly classified as normal.

### C. Limitations

The MIT-BIH arrhythmia dataset does have some severe class imbalance, and therefore some classes have very few samples with which models can properly learn differences in morphology, shown in Figure 1. The data also comes from very few patients, and therefore does not integrate a diverse set of patient characteristics like the datasets used by Hannun et al. and Ribeiro et al. in [8] and [9], which are shown to be very generalizable due to the large datasets used to train these models.

Secondly, this data was recorded between 1975 and 1979. While the morphology of an ECG taken from a human heart is unlikely to have changed since then, software capabilities (specifically methods to prevent noise) will likely have improved the quality of ECGs being recorded in modern machines.

While our method could be used to highlight some of these issues, for example allowing us to speculate a particular segment is being highlighted due to confounding factors, a larger, more recent dataset of ECGs taken from patients with a diverse set of characteristics is needed.

## V. CONCLUSION

In this study, we have proposed a tool for addressing the interpretability issue of ML ECG classification. We first build two models based on state-of-the-art architectures, a CNN and an LSTM. The MIT-BIH arrhythmia dataset is then split into beats classified into eight classifications. Saliency maps are used to visualize what areas of each beat are important to the final classification decision made by the model. Finally, we segment each saliency map into blocks, repeated for each beat every class, in order to quantitatively compare what areas the model is looking at for each class, and investigate any differences.

Model interpretability is necessary when applying these models to clinical applications. Our method provides a technique for quantitatively comparing which section of an ECG a model is highlighting for each class. Unlike comparing per-ECG saliency maps, this allows us to determine the patterns which are being picked up by the model for each class as a whole, allowing us to compare this to clinical understanding and investigate any outliers. The process allows us to determine where confounding variables such as noise or patient characteristics are being picked up inappropriately, and where the model is picking up new and unexpected features which could offer new insights to prediction models in the future.

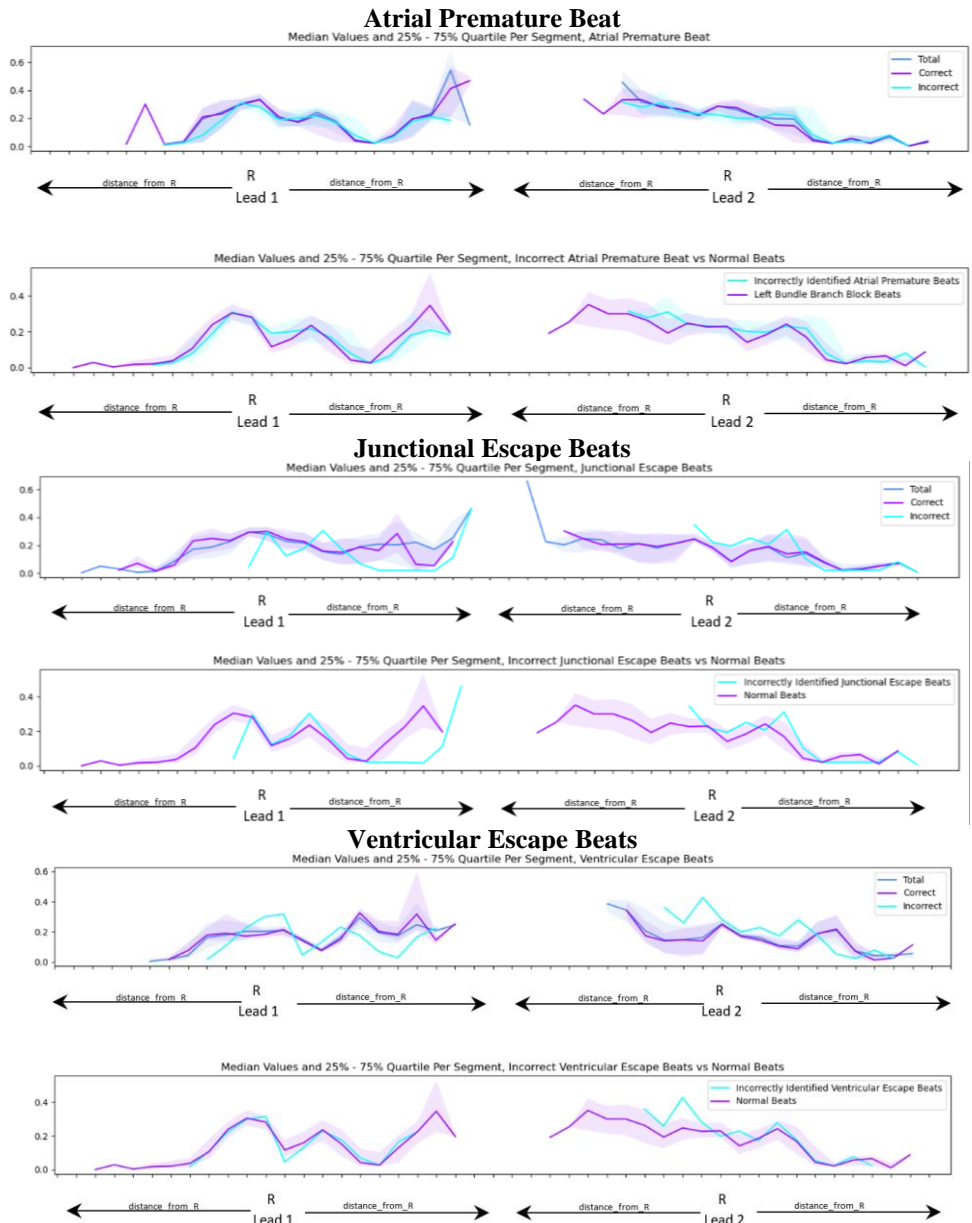

*Figure 8: Saliency Segment Analysis of Incorrectly Identified Beats*